\documentstyle[12pt]{article}
\begin{document} 
\vspace*{-1in} 
\renewcommand{\thefootnote}{\fnsymbol{footnote}} 
\begin{flushright} 
TIFR/TH/98-48\\
November 1998\\ 
hep-ph/9811501 
\end{flushright} 
\vskip 65pt 
\begin{center} 
{\Large \bf Getting to the top with extra dimensions}\\
\vspace{8mm} 
{\bf 
Prakash Mathews${}^{1}$\footnote{prakash@theory.tifr.res.in}, 
Sreerup Raychaudhuri${}^2$\footnote{sreerup@hecr.tifr.res.in},   
K.~Sridhar${}^1$\footnote{sridhar@theory.tifr.res.in}
}\\ 
\vspace{10pt} 
{\sf 1) Department of Theoretical Physics, Tata Institute of 
Fundamental Research,\\  
Homi Bhabha Road, Bombay 400 005, India. 

2) Department of High Energy Physics, 
Tata Institute of Fundamental Research, \\  
Homi Bhabha Road, Bombay 400 005, India. } 
 
\vspace{80pt} 
{\bf ABSTRACT} 
\end{center} 
\vskip12pt 
The prospect of large extra dimensions and an effective theory
of gravity at around a TeV has interesting experimental consequences.
In these models, the Kaluza-Klein modes interact with Standard
Model particles and these interactions lead to
testable predictions at present and planned colliders. We investigate
the effect of virtual exchanges of the spin-2 Kaluza-Klein modes in the
production cross-section of $t \bar t$ pairs at the Tevatron and
the LHC and find that the $t \bar t$ cross-section can be an
effective probe of the large extra dimensions. This enables us
to put bounds on the effective low-energy scale.
 
\setcounter{footnote}{0} 
\renewcommand{\thefootnote}{\arabic{footnote}} 
 
\vfill 
\clearpage 
\setcounter{page}{1} 
\pagestyle{plain}
While traditionally it has been assumed that quantum gravity
effects are large only near the Planck scale, $M_P=1.2 \times
10^{19}$~GeV, recently it is becoming evident \cite{anto, dimo}
that the effects of gravity propagating in dimensions higher than
four can completely change this simple picture, if these higher
dimensions are compactified to have relatively large sizes.
The proposal finds its most natural setting in a string-theoretic
framework: the feature of large extra dimensions is quite
generic and can be realised in the context of several string
models \cite{string}. One starts with a theory in $D$ dimensions and
the extra dimensions are
then compactified to obtain the effective low-energy theory in
3+1 dimensions, and it is assumed that $n$ of these extra dimensions
are compactified to a common scale $R$ which is relatively large,
while the remaining dimensions are compactified to much smaller
length scales which are of the order of the Planck scale. One immediate
advantage of the string construction is that matter is localised on
a 3-brane because the
open strings, to which correspond the SM particles, end on the brane
and are therefore confined to the $3+1$-dimensional spacetime, while
gravitons (corresponding to closed strings) propagate in the 
$4+n$-dimensional bulk \cite{string, sundrum}. The relation between 
the scales in $4+n$ dimensions and in $4$ dimensions is given by \cite{dimo}
\begin{equation} 
M^2_{\rm Pl}=M_{S}^{n+2} R^n ~,
\label{e1} 
\end{equation} 
where $M_S$ is the low-energy effective string scale. Then it follows that 
if we need $M_S$ to be of the order of a TeV then $R=10^{32/n -19}$~m.
One immediate consequence of this relation is that for this value of $R$ 
(for any given $n$) we will expect to see deviations from Newton's law
of gravitation. The case $n=1$ is obviously excluded, but for $n=2$ or
larger it is possible to have the scale $M_S$ to be of the order of 1~TeV
and still be consistent with experiment. For $n=2$ the comapctified dimensions
are of the order of 1 mm and just below the experimentally tested region
for the validity of Newton's law of gravitation, and within the possible
reach of ongoing experiments \cite{gravexp}.

For particle physics, one very important consequence of the lowering of the
string scale is the nullification of the hierarchy problem. This is because
the string scale is now of the order of a TeV and of the same order as the 
electroweak scale. In fact, it has been shown \cite{dimo2} that is 
possible to construct a phenomenologically viable scenario with large
extra dimensions, which can survive the existing astrophysical and 
cosmological constraints.
Recently, the effect of the Kaluza-Klein states on the
running of the gauge couplings i.e. the effect of these states on the beta
functions of the theory have been studied \cite{dienes} and it has been 
shown that the GUT scale can be also lowered down to scales close to 
the electroweak scale.

Let us describe the effective theory below the scale $M_S$ that emerges
in such a scenario \cite{sundrum, grw, hlz}.
The particle content of the theory below the scale
$M_S$ is as follows~: there are the usual Standard Model (SM) particles 
and the graviton and also other light modes related to the brane dynamics.
In particular, there are the $Y$ modes which are related to the
deformation of the brane and these are the massless Nambu-Goldstone
bosons (in the higher-dimensional theory) due to the spontaneous breaking 
of translational 
invariance in the transverse directions. In the effective theory, however, 
some of these modes could acquire mass \cite{dimo2}.
In any case, the $Y$ modes couple to the SM matter
only in pairs so their effects, compared to the graviton, are subleading.
The graviton corresponds to a tower of Kaluza-Klein modes
which contain spin-2, spin-1 and spin-0 excitations. The spin-1
modes do not couple to the energy-momentum tensor and their
couplings to the SM particles in the low-energy effective
theory is not important. The scalar modes couple to the trace
of the energy-momentum tensor, so they do not couple to massless
particles. 

The interesting aspect of the large extra dimensions is that it is
possible to have experimentally observable consequences of the effects of the
infinite tower of Kaluza-Klein states in present and planned high-energy
collider experiments. As explained above, the effects of the spin-1
Kaluza-Klein states can be neglected. Further for collider experiments,
it is usual to consider the particles in the initial state as massless. In
such a case, the scalar contributions can also be neglected. The
only states that contribute are the spin-2 Kaluza Klein states. These
correspond to a massless graviton in the $4+n$ dimensional theory,
but manifest as an infinite tower of massive gravitons in the low-energy
effective theory. For graviton momenta smaller than the scale $M_S$, the
effective description reduces to one where the gravitons in the bulk 
propagate in the flat background and couple to the SM fields which live
on the brane via a (four-dimensional) induced metric $g_{\mu \nu}$. 
Moreover, it turns out that the couplings of the modes in the
Kaluza-Klein expansion of the energy-momentum tensor, $T_{\mu \nu}$,
to the SM fields are universal. This model-independence of the couplings
is crucial because it allows us to make definite predictions for the
experimental consequences of these interactions.

With the above assumptions and starting from a linearized gravity Lagrangian
in $D$ dimensions, the four-dimensional interactions can be derived after
a Kaluza-Klein reduction (on a $n$-dimensional torus) has been performed.
The interaction of the SM particles with the graviton can be derived from 
the following Lagrangian:
\begin{equation} 
{\cal L}=-{1 \over \bar M_P} G_{\mu \nu}^{(j)}T^{\mu\nu} ~,
\label{e2} 
\end{equation} 
where $j$ labels the Kaluza-Klein mode and $\bar M_P=M_P/\sqrt{8\pi}$.
Using the above interaction Lagrangian the couplings of the graviton
modes to the SM particles can be calculated \cite{grw,hlz}.

Very recently, there have been papers which have studied the
consequences at colliders of this TeV scale effective theory
of gravity. In particular, direct searches for graviton production
at $e^+ e^-$ and $p \bar p$ and $pp$ colliders have been suggested 
\cite{grw, mpp, hlz}. These lead to spectacular single photon + missing 
energy or monojet + missing energy signatures. Constraints coming
from indirect effects (i.e. the study of the effects of virtual gravitons
in various experimental observables) can also be made. The virtual effects
in $e^+ e^- \rightarrow f \bar f$ and in high-mass dilepton production
at Tevatron and LHC have been studied \cite{hewett}. In view of the
fact that the effective Lagrangian given in Eq.~\ref{e2} is suppressed
by $1/\bar M_p$, it may seem that the effects at colliders will be hopelessly
suppressed. However, in the case of real graviton production, the phase
space for the Kaluza-Klein modes cancels the dependence on $\bar M_p$ 
and, instead, provides a suppression of the order of $M_S$. For the
case of virtual production, we have to sum over the whole tower of 
Kaluza-Klein states and this sum when properly evaluated \cite{hlz, grw}
again substitutes the scale $M_S$ for $\bar M_p$.

In the present work, we study the effect of the virtual graviton
exchange on the production cross-section of the $t \bar t$ at hadron 
colliders. In addition to the SM cross-section, we have new $s$-channel
production mechanisms for $t \bar t$ production where the 
graviton modes couple to the $q \bar q$ or $gg$ initial state. Given the
vertices for the $q \bar q G^{(j)}$ coupling and the $g g G^{(j)}$
coupling and summing over all the graviton modes, as discussed explicitly
in Refs.~\cite{hlz, grw}, we find the following expressions for the
cross-sections involving the virtual graviton exchange:
\begin{eqnarray} 
&&{d\hat \sigma \over d\hat t}(q \bar q \rightarrow t \bar t) 
= {d\sigma \over d\hat t}_{\rm SM} (q \bar q \rightarrow t \bar t) 
 + {\pi \lambda^2 \over 64 \hat s^2 M_S^8} \nonumber \\
&&\biggl \lbrack 5 \hat s^2(\hat t- \hat u)^2+4 (\hat t-\hat u)^4 
+8 \hat s (\hat t- \hat u)^2 (\hat t+\hat u)-2 \hat s^3 (\hat t+\hat u) -\hat 
s^4\biggr\rbrack ~,
\label{e3} 
\end{eqnarray} 
and
\begin{eqnarray} 
\!\!\!&&{d\hat \sigma \over d\hat t}(gg \rightarrow t \bar t) 
= {d\sigma \over d\hat t}_{\rm SM} (gg \rightarrow t \bar t) 
 - {\pi \over 16 \hat s^2 } \biggl \lbrack {3 \lambda^2 \over M_S^8} 
- {2\alpha_s \over M_S^4}{\lambda \over (M_t^2-\hat t)(M_t^2-\hat u)} 
\biggr\rbrack \nonumber \\
\!\!\!&&\quad \times\biggl\lbrack 6 M_t^8-4 M_t^6(\hat t+\hat u) 
+4M_t^2 \hat t \hat u 
(\hat t+ \hat u) -\hat t\hat u(\hat t^2 + \hat u^2) + M_t^4(\hat t^2 -6 
\hat t\hat u +\hat u^2)\biggr\rbrack ~.
\label{e4} 
\end{eqnarray} 
In the above $M_t$ refers to the mass of the top quark and the
coupling $\lambda$ is the effective coupling at the scale $M_S$.
$\lambda$ is expected to be of ${\cal O}(1)$, but its sign
is not known $a\ priori$. In our work we will explore the sensitivity
of our results to the choice of the sign of $\lambda$.
There is no interference between the SM diagram and the graviton
exchange diagram in the $q \bar q$-initiated cross-section, but there
is a non-vanishing interference in the $gg$-initiated cross-section.
With the above expressions for the subprocess cross-sections at hand, it
is staighforward to compute the integrated top-quark cross-section,
using the formula
\begin{equation}
\sigma (AB \rightarrow t \bar t)= \sum \int dx_1
dx_2 d\hat t ~\lbrack f_{a/A}(x_1)~ f_{b/B}(x_2) + x_1 \leftrightarrow x_2 
\rbrack ~{d\hat \sigma \over d\hat t} ~,
\label{e5}
\end{equation}
where $A,\ B$ are the initial hadrons (either $p \bar p$ or $pp$), and
$f_{k/h}$ denotes the probability of finding a parton $k$ in the hadron $h$.
The sum in Eq.~\ref{e5} runs over the contributing subprocesses.

Before we discuss the results of our computation,
a few remarks about the significance of higher-order corrections
are in order. The cross-sections presented above are at the lowest
order in perturbation theory. In the QCD case,
significant progress has been made in computing higher-order
corrections to heavy quark production. Not only have the 
next-to-leading order corrections been calculated a long time 
ago \cite{nason}, but the resummation of soft gluons and 
its effect on the total cross-section have been recently 
computed \cite{berger, catani}. In principle, a reliable estimate 
of the cross-section for the case under consideration
can also be made only when we have 
(at least) the corrections to these processes at next-to-leading 
order. But for want of such a calculation, the best we can
do is to use the leading order QCD and the resummed QCD
cross-sections \cite{berger} to extract a `$K$-factor'. We
note that the resummed cross-sections of Ref.~\cite{catani}
would yield a different $K$-factor, but our bounds on the
new physics scale are not affected seriously by this change.
We work with the approximation that the new physics
will also be affected by QCD corrections in a similar fashion so that
we can fold in our cross-sections for this case by the 
same $K$-factor. Clearly, more work is needed in this direction
but we do not expect that our results will be qualitatively changed by
higher order QCD corrections.

\begin{figure}[h]
\begin{center}
\vspace*{5.0in}
      \relax\noindent\hskip -5.2in\relax{\includegraphics{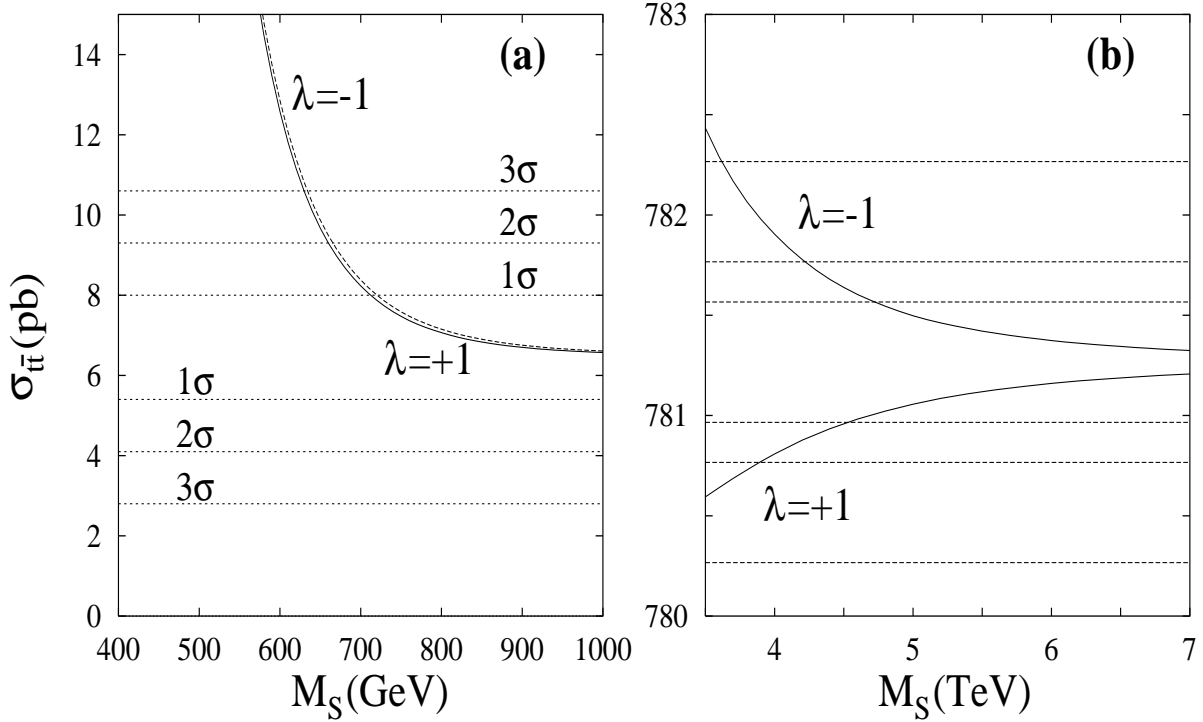}}
\end{center}
\vspace{-1.6in}
\caption{\footnotesize\it Illustrating the variation of the $t \bar t$
                          cross-section with variation in the scale
                          $M_S$ at (a) the Tevatron and (b) the LHC.
                          For the Tevatron, dashed lines show the 
                          experimental data from Run I, when the CDF
                          and D0 results are combined. For the LHC,
                          the dashed lines correspond to errors on
                          the SM cross-section of 0.3, 0.5 and 1 pb 
                          respectively. } 
\end{figure}

We present the results of our numerical computations in Fig.~1. In
Fig~1 (a), we have plotted the cross-section as a function of the scale $M_S$
for $p \bar p$ collisions at the Tevatron energy of $\sqrt{s}=1.8$~TeV. 
We have used the CTEQ4M densities \cite{cteq} and the parton distributions
are taken from PDFLIB \cite{pdflib}. 
As explained earlier, we have set the magnitude of the coupling 
$\lambda=1$, but we still have the freedom to choose its sign. We have
plotted the curves for both choices of sign. We find that the cross-section
for the case $\lambda=-1$ is larger than the case $\lambda=1$. This is
due to the fact that in the former case the interference contribution
in the $gg$-initiated channel makes a positive contribution. For the
experimental value of the cross-section we have used the CDF-D0 average
given in Ref.~\cite{tevatron}. The $1\sigma$, $2\sigma$, $3\sigma$ bands
of this average cross-section is shown in Fig~1 (a). We see that for the
case $\lambda=-1$, a bound of about 665~GeV results at the 95\% confidence
level. In the $\lambda=1$ case, this turns out to be about 660~GeV.
These bounds are expected to get much better at Tevatron Run II.
We expect (assuming an integrated luminosity of 2 fb) at the Tevatron
the $2\sigma$ bound to be about 800~GeV for $\lambda=\pm 1$. 

Going from Tevatron to LHC ($pp$ collisions at $\sqrt{s}=14$~TeV), 
does affect the results quite significantly.
Because of the dominant $gg$ channel at the LHC energy we find that
the effect of the interference term is seen much more dramatically in the
results, when $\lambda$ changes from +1 to -1 (see Fig.~1 (b)). 
The value of the cross-section at the LHC energy is large (about
868 pb) and with an expected luminosity of 10~fb${}^{-1}$, we expect of
the order of
$8.6\times 10^{6}$ events at LHC. The statistical error is consequently 
negligibly small and the error is expected to be dominated by the
systematics. We assume errors of the order of
0.3, 0.5 and 1.0 pb (where 0.3~pb is the $1\sigma$ statistical error)
and with these assumptions, we find that at LHC we can
probe the effective string scale to mass values of the order of 2.5--5~TeV. 

In summary, recent proposals about strong effects of gravity at TeV
scale can be tested at colliders. We have probed the effects of the
interactions of the spin-2 Kaluza-Klein modes with SM matter in the
production of $t \bar t$ pairs at the Tevatron and LHC. We find that
this process can be a useful channel to put bounds on the effective
theory scale $M_S$ by using the experimentally measured $t \bar t$
cross-sections. The bounds from Tevatron are around 665~GeV at the
95\% confidence level for $\lambda=-1$ and about 660~GeV for $\lambda=1$.
More accurate measurements of the top production cross-section at
Tevatron Run II will improve these bounds by about 25~GeV. At LHC, it is 
expected that $M_S$ values of about 2.5--5~TeV can be probed in this channel.

\end{document}